\documentstyle[preprint,aps,prd,epsfig,amssymb,psfrag,rotate]{revtex}

\newcommand{\ovl}{\overline}
\newcommand{\wt}{\widetilde}
\newcommand{\bea}{\begin{eqnarray}}
\newcommand{\eea}{\end{eqnarray}}
\newcommand{\nl}{\nonumber\\}
\newcommand{\al}{\alpha}
\newcommand{\be}{\beta}
\newcommand{\ga}{\gamma}
\newcommand{\de}{\delta}
\newcommand{\Ga}{\Gamma}
\newcommand{\si}{\sigma}

\newcommand{\xw}{\sin^2\theta_W}

\begin{document}
\draft
\preprint{\begin{tabular}{l}
\hbox to\hsize{hep-ph/0205013 \hfill KIAS-P02034}\\[5mm] \end{tabular} }

\title{ 
Very light sbottom and gluino scenario confronting
electroweak precision tests
} 
\author{Seungwon Baek}
\address{ 
School of Physics, KIAS, 207-43 Cheongryangri-dong, Seoul 130-012, Korea}

\date{\today}

\maketitle
\thispagestyle{empty}

\begin{abstract}
The minimal supersymmetric model with  light sbottom ($\sim m_b$) and
light gluino (12 $\sim$ 16 GeV) can explain the excess of bottom quark
production cross section at hadron colliders and has drawn much
attention.  We calculate one-loop contribution to Z-pole observable
$R_b$ in this scenario when large CP violating phases are allowed in
the gluino mass and/or sbottom sector.  We show that these large CP
violating phases can suppress the new contribution $\delta R_b$ safely
within the experimental bounds even for heavier sbottom too heavy
($\wt{b}_2 \gtrsim$ 200 GeV) to be produced at LEP2 in association
with light sbottom $\wt{b}_1$.
\end{abstract}

\pacs{PACS numbers:}



\section{Introduction}
The minimal supersymmetric standard model (MSSM) is the most promising
candidate for the physics beyond the standard model (SM).  In its unconstrained
version, it has many free parameters, $\sim {\cal O}(100)$, 
even in the R-parity
conserving scenario. These free parameters will be ultimately determined from
both theoretical constraints and experimental searches.

The negative searches of sparticles at the colliders
put strong lower bounds on their masses~\cite{Groom:in}. 
However, the bounds are usually derived from some model-dependent assumptions
and they are not applicable if these assumptions are relaxed.
Actually, there has been much interest in the possibility that 
light sbottom ($\sim m_b$) and light gluino ($\sim$ 12 to 16 GeV)
may have escaped the direct searches~\cite{Dedes:2000nv,Berger:2000mp}.

The MSSM scenario with light sbottom and light gluino can explain
the discrepancy between the SM prediction and the data from hadron colliders 
in the bottom-quark production.
To fit the data it is necessary to
have light gluino ($\sim$ 12 to 16 GeV) which decays with 
100\% branching fraction into
bottom quark $b$ and a light bottom squark 
($m_{\wt{b}_1} \sim $ 2 to 5.5 GeV)~\cite{Berger:2000mp}.

If the light sbottom, $\wt{b}_1$, couples strongly with $Z$ boson, it can give
large contribution to $Z$ peak observables. However, the mixing angle in the
sbottom sector can be chosen in such a way that $Z\wt{b}_1 \wt{b}_1$ couplings
are small~\cite{Carena:2000ka}.
However, this tuning is not enough to suppress all the Z-pole observables.
Recently, J. Cao and Z. Xiong and J. M. Yang~\cite{Cao:2001rz} 
showed that there can be
large contribution to $Zb\ovl{b}$ vertex through gluino-sbottom
exchanged one-loop
diagrams in the CP conserving MSSM.

In general there can be large number of CP violating phases in the MSSM 
Lagrangian, and a natural expectation on the size of the phases
is ${\cal O}(1)$.
However, most of them should be suppressed $\lesssim 10^{-2}$ to satisfy
the bound on the electric dipole moments (EDM) of neutron and electron. 
On the contrary, it is known that ${\cal O}(1)$ CPV phases other than
CKM phase in the SM are necessary
to explain the asymmetry between the matter and antimatter in the 
universe~\cite{Cline:1998hy}.
This is the SUSY CP problem. There are some suggestions to solve this problem,
for example, decoupling solution~\cite{decoupling}.
Here we implicitly take one of the solutions of SUSY CP problem to
allow ${\cal O}(1)$ phase while satisfying the EDM bounds.

Specifically we consider the effects of phase $\varphi_3$ on the 
gluino mass parameter $M_3$ and a phase $\varphi_b$ in the 
down squark mass-squared matrix. We show that the effect of 
the CP violating (CPV)
phases, $\varphi_3,\varphi_b$, on the $Zb\ovl{b}$ in the light sbottom/gluino
scenario is nontrivial and can actually alleviate the strong Z-peak constraints.

This paper is organized as follows.
In Section II we present our result on the one-loop calculation of
SUSY QCD correction to $Zb\ovl{b}$ vertex. The numerical analysis
is done in Section III. In Section IV we conclude.

\section{The effect of $\phi_3$ and $\phi_b$ on $R_b$}

The gluino mass parameter in the soft SUSY breaking Lagrangian
is a complex parameter, 
$M_3 \equiv |M_3| e^{i \varphi_3} = m_{\tilde{g}} e^{i \varphi_3}$. 
The sbottom mass-squared matrix
$\wt{m}^2_b$ also has a CPV phase $\varphi_b \equiv \arg(A_b^* -\mu \tan\beta)$, 
where $A_b$
is a trilinear coupling in the soft supersymmetry breaking terms
and $\mu$ is the Higgs mixing
term in the superpotential. The sbottom mass-squared matrix $\wt{m}^2_b$ is
diagonalized by a unitary matrix $\Gamma_b$ as
\begin{eqnarray}
\Gamma_b  \wt{m}^2_b \Gamma_b^\dagger = 
   {\rm diag} (m^2_{\wt{b}_1},m^2_{\wt{b}_2}),
   \quad m^2_{\wt{b}_1}<m^2_{\wt{b}_2}
\end{eqnarray}
where $\Gamma_b$ can be parametrized as
\bea
  \Gamma_b = \left(
      \begin{array}{cc}
         c_b & e^{i \varphi_b} s_b \\
         e^{-i \varphi_b} s_b & c_b 
      \end{array}
    \right), \qquad \mbox{$c_b \equiv \cos\theta_b,\;s_b \equiv \sin\theta_b$}.
\eea
We also introduce the following notation for later use,
\bea
   \Gamma^b_L \equiv \left(c_b, e^{-i \varphi_b}  s_b\right)^T, \quad
   \Gamma^b_R \equiv \left(e^{i \varphi_b} s_b,  c_b \right)^T.
\eea

We note that $Z\wt{b}_1 \wt{b}_1 \propto -1/2 c_b^2 +1/3 \xw$ 
coupling is independent of CPV phases
and vanishes by choosing $|c_b|=\sqrt{2/3} \sin\theta_W \approx 0.39$. 
We will assume
$|c_b| \approx$ 0.3 -- 0.45 to
suppress the tree-level contribution to $R_b$~\cite{Carena:2000ka}. 
In this case, however,
$Z\wt{b}_1 \wt{b}_2 \propto e^{i \varphi_b} c_b s_b$ can be sizable.
Since LEP2 did not see an event like $e^+e^- \to \wt{b}_1 \wt{b}_2$
it gives a constraint on the heavy sbottom 
mass to be $m_{\tilde{b}_2} \gtrsim 200$ GeV.

\begin{figure}[thb]

\centerline{\epsfxsize=16.cm \epsfbox{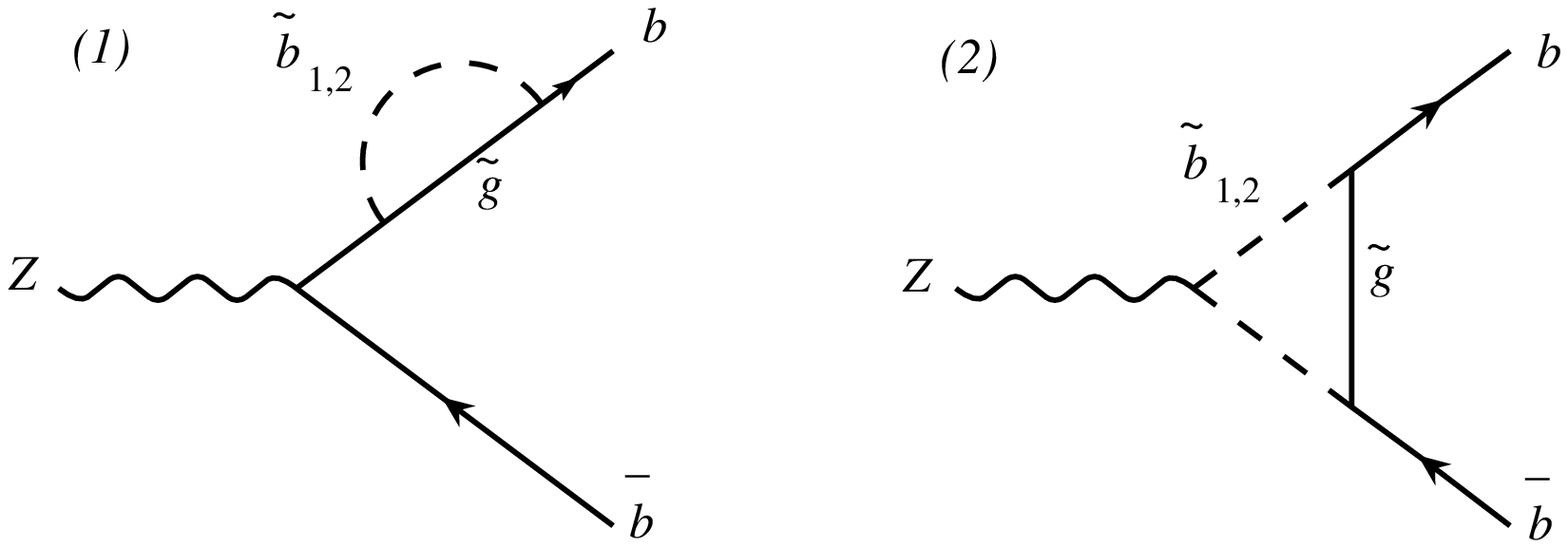}}
\caption{Feynman diagrams with sbottom-gluino loop which contribute to
the correction on $Zb\ovl{b}$ vertex}
\label{Zbb_1.ps}
\end{figure}

In Fig.~\ref{Zbb_1.ps}, we show the Feynman diagrams with
sbottom-gluino loop which contribute to
the correction on $Zb\ovl{b}$ vertex. The diagrams diverge and we used
dimensional regularization and on-shell renormalization scheme to calculate them.
After including the one-loop contributions, the effective $Zb\ovl{b}$-vertex
can be written as~\cite{Cao:2001rz}
\bea
\label{eq:Zbb}
  V^{\rm eff}_\mu(Zb\ovl{b}) &=& -i g_Z \Big[
   \ga_\mu (g_L^b P_L + g_R^b P_R) \nl
 &&    +{\al_s \over 3 \pi} 
     \left(F_L^V \ga_\mu P_L +F_R^V \ga_\mu P_R
       +F_L^T i \si_{\mu\nu} k^\nu P_L +F_R^T i \si_{\mu\nu} k^\nu P_R
     \right)
   \Big],
\eea
where $g_Z =g/\cos\theta_W$, $g_L^b=-1/2+\xw/3$, $g_R^b=\xw/3$ and
$k$ is $Z$-boson momentum.
The form factors are expressed in terms of one-loop two-point and three-point
functions
\bea
\label{eq:ffV}
  F_L^V &=& 2 g_L^b \sum_{A=1,2} \Big[ (G_{LL})_{AA}
       \left(B_1(A) + 2 m_b^2 {\partial B_1(A) \over \partial p_b^2}\right) 
          + 2 m_{\tilde{g}} m_b (G_{RL})_{AA}
                      {\partial B_0(A) \over \partial p_b^2}
 \Big]_{p_b^2=m_b^2}   \nl
  && +2 \sum_{A,B=1,2} 
  \left(-{1 \over 2} \wt{K}^{bb}_{AB} + {1 \over 3} \xw \delta_{AB} \right)
 \Bigg\{
   2 (G_{LL})_{AB} C_{00}(A,B) \nl
 && + m_{\tilde{g}} m_b (G_{RL}+G_{LR})_{AB}
         \Big(C_0(A,B)+C_1(A,B)+C_2(A,B)\Big) \nl
 &&   + m_b^2  (G_{LL}+G_{RR})_{AB}
    \Big(C_1(A,B)+C_2(A,B)+C_{11}(A,B)+C_{22}(A,B)+2 C_{12}(A,B)\Big)
 \Bigg\} \nl
 F_L^T &=& -2 \sum_{A,B=1,2} 
\left(-{1 \over 2} \wt{K}^{bb}_{AB} + {1 \over 3} \xw \delta_{AB} \right) 
\Bigg\{
m_{\tilde{g}} (G_{RL})_{AB} \Big(C_0(A,B)+C_1(A,B)+C_2(A,B)\Big) \nl
&&+m_b(G_{LL})_{AB}  \Big(C_1(A,B)+C_{11}(A,B)+C_{12}(A,B)\Big) \nl
&& +m_b(G_{RR})_{AB}  \Big(C_2(A,B)+C_{22}(A,B)+C_{12}(A,B)\Big) \Bigg\},
\eea
where $\wt{K}^{bb}_{AB} \equiv (\Ga_L^b)_{A1} (\Ga_L^b)_{B1}^*$,
$(G_{LL})_{AB} \equiv(\Ga_L^b)_{A1}^* (\Ga_L^b)_{B1}$,
 $(G_{RR})_{AB} \equiv (\Ga_R^b)_{A1}^* (\Ga_R^b)_{B1}$,
 $(G_{LR})_{AB} \equiv e^{i\varphi_3}(\Ga_L^b)_{A1}^* (\Ga_R^b)_{B1}$ and
 $(G_{RL})_{AB} \equiv e^{-i\varphi_3}(\Ga_R^b)_{A1}^* (\Ga_L^b)_{B1}$.
The chirality flipped form factors $F_R^V$ and $F_R^T$ are obtained by
interchanging $L$ with $R$ in (\ref{eq:ffV}).
For the loop functions we followed the convention in the 
Ref.~\cite{Hahn:1998yk},
\bea
 B_0(A,B) &=& B_0(m_b^2,m_{\tilde{g}}^2,m^2_{\tilde{b}_A}), \nl
 C_0(A,B) &=& 
  C_0(m_b^2,m_Z^2,m_b^2,m_{\tilde{g}}^2,m^2_{\tilde{b}_A},m^2_{\tilde{b}_B}), 
\eea
{\it etc}.

The Z-pole observable $R_b=\Ga(Z\to b\ovl{b})/\Ga(Z\to \mbox{hadrons})$
is precisely measured and agrees well with the SM prediction and
therefore gives a stringent constraint on some new physics models.
In the model at hand $R_b$ is given by
\bea
  \de R_b &\equiv& R_b -R_b^{\rm SM} \nl
  &=& R_b^{\rm SM} (1-R_b^{\rm SM}) {2 \al_s \over 3 \pi}
    \frac{1}
{(3+\be^2)\left(\left(g_L^{b}\right)^2+\left(g_R^{b}\right)^2\right)
  +6(1-\be^2) g_L^b g_R^b} \nl
 &&\times \Bigg[  (3+\be^2) (g_L^b {\rm Re} F_L^V+ g_R^b {\rm Re} F_R^V ) 
  +3(1-\be^2) (g_L^b {\rm Re} F_R^V+ g_R^b {\rm Re} F_L^V ) \nl
&&  +6 m_b (g_L^b+g_R^b) ({\rm Re} F_L^T+  {\rm Re} F_R^T ) \Bigg],
\eea
where $\be=\sqrt{1-4 m_b^2/m_Z^2}$.
In \cite{Cao:2001rz} it was shown that $R_b$ is actually a severe constraint
on the CP conserving MSSM with very light gluino and sbottom. They showed that 
the contributions of light sbottom and heavy sbottom to $R_b$ tend to
cancel each other if the heavier sbottom is 
light enough $( \lesssim 200 {\rm GeV})$.
Since this light $\wt{b}_2$ is unlikely by LEP2 searches as mentioned
above, they argued that this SUSY scenario is disfavored.
We will show that $\wt{b}_2$ can be heavy enough to escape the LEP2 search
while satisfying the $R_b$ constraint in the CPV MSSM scenario, which
is the topic of next section.

\section{Numerical Analysis}
In this section we will show that the severe constraint on the 
light sbottom/gluino
in the CP conserving SUSY model can be alleviated if MSSM has additional 
CP violating phases other than the CKM phase in the SM. For this analysis we take
the following values for the input parameters~\cite{Drees:2001xw},
\bea
  \al_s = 0.1192, \quad
  m_b = 4.8 \;{\rm GeV}, \quad
  m_Z = 91.188 \;{\rm GeV}, \quad
  \xw = 0.2312.
\eea
We also take $m_{\tilde{g}}=14$ GeV and $m_{\tilde{b}_1}=5$ GeV. 
We have checked that
the  variation of $m_{\tilde{g}} (m_{\tilde{b}_1})$ in the range 12--16 GeV
(2--5.5 GeV) does not significantly change our results.
We vary the phases $\varphi_{3}$ and $\varphi_b$ from 0 to $2 \pi$.

\begin{figure}[thb]
\centerline{\epsfxsize=10.cm \epsfbox{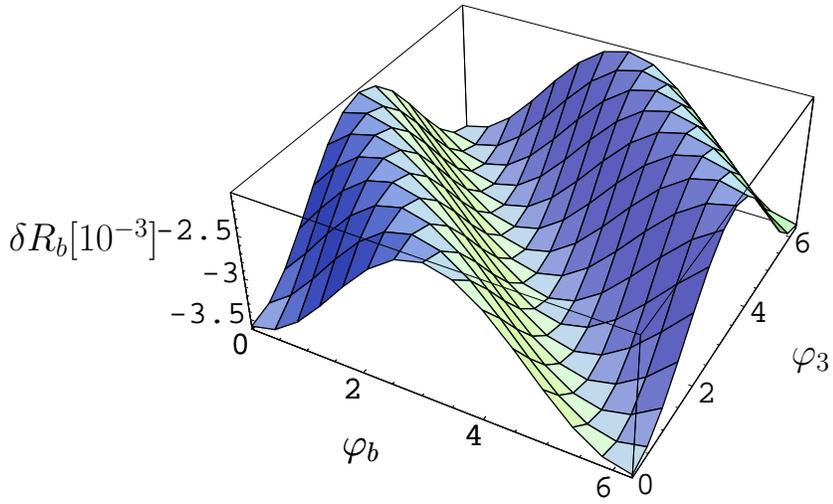}}
\caption{$\de R_b$ as a function of $\varphi_b$ and $\varphi_3$}
\label{Rb-phi}
\end{figure}

In Fig.~\ref{Rb-phi} $R_b$ is shown as a funtion of $\phi_b$ and $\phi_3$
for $c_b = 0.39$ and $m_{\tilde{b}_2}=200$ GeV.
We can see that $\de R_b$ is maximized in the CP conserving case and reduced
significantly when the CPV phases $\phi_3$ and $\phi_b$ are turned on.
$\de R_b$ is minimized when $\phi_b +\phi_3 \approx \pi$.

\begin{figure}[thb]
\centerline{\epsfxsize=10.cm \epsfbox{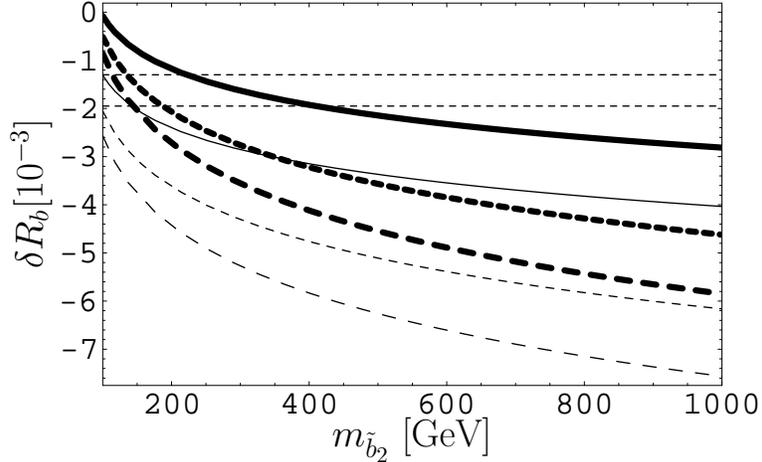}}
\caption{Thick (thin) lines represent $\de R_b$ as a function of 
$m_{\tilde{b}_2}$ for $c_b=0.3,0.39,0.45$ (from above) in the CP violating
(conserving) MSSM. The dashed horizontal lines
represent 2$\sigma$ and 3$\sigma$ (from above) deviations from the experimental
central value.}
\label{Rb-msb2}
\end{figure}

In Fig.~\ref{Rb-msb2} we show $\de R_b$ as a function of $m_{\tilde{b}_2}$
for $c_b=0.3,0.39,0.45$ (from above). 
We got similar results with \cite{Cao:2001rz} in the CP conserving limit (thin
lines). 
In the CPV case we took $\varphi_3 =\varphi_b =\pi/2$ (thick lines) which 
is a maximal CP violating case. We can see that 
the heavier sbottom mass can be pushed
beyond the LEP2 search if $c_b <0.39$ while keeping $R_b$ within
the experimental value at the 3-$\sigma$ level. 
Note that the SM precition $R_b^{\rm SM}=0.21596$ differs from
the experimental value $R_b^{\rm exp} = 0.21646 \pm 0.00065$~\cite{Drees:2001xw} 
by 0.8$\sigma$
which can further reduce the deviation of the SUSY QCD contribution to
$R_b$ from the measurement.

\begin{figure}[thb]
\centerline{
\begin{tabular}{cc}
{\epsfxsize=7.cm \epsfbox{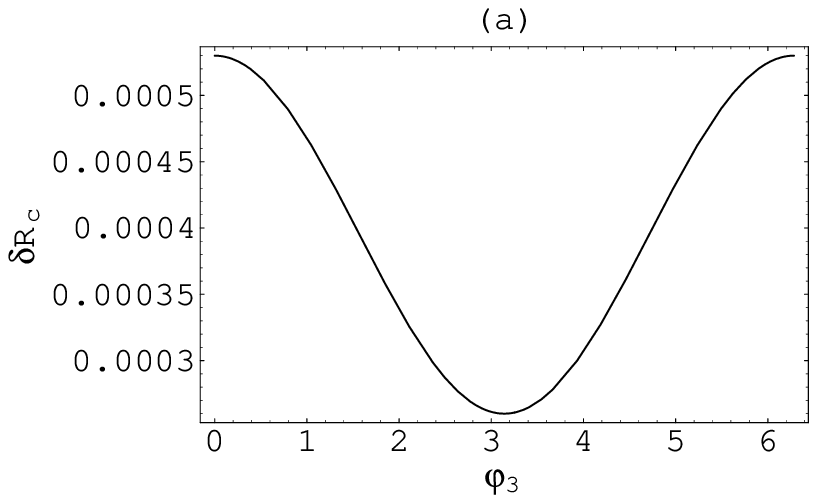}} &
{\epsfxsize=7.cm \epsfbox{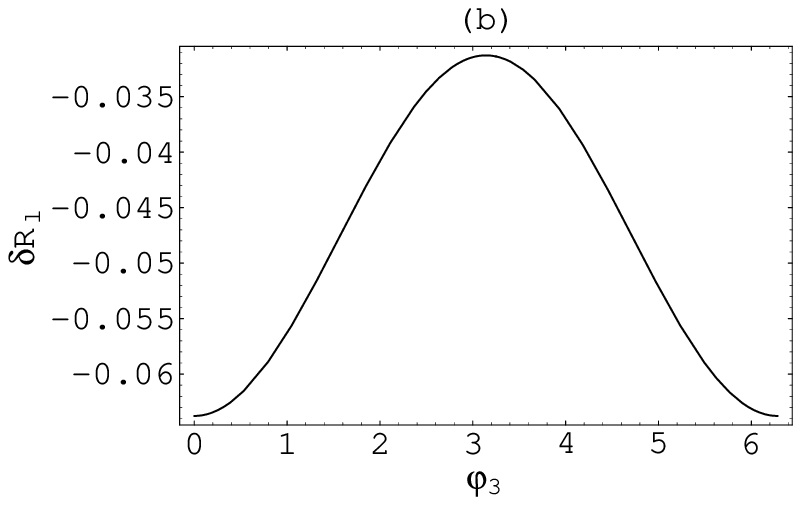}}\\
{\epsfxsize=7.cm \epsfbox{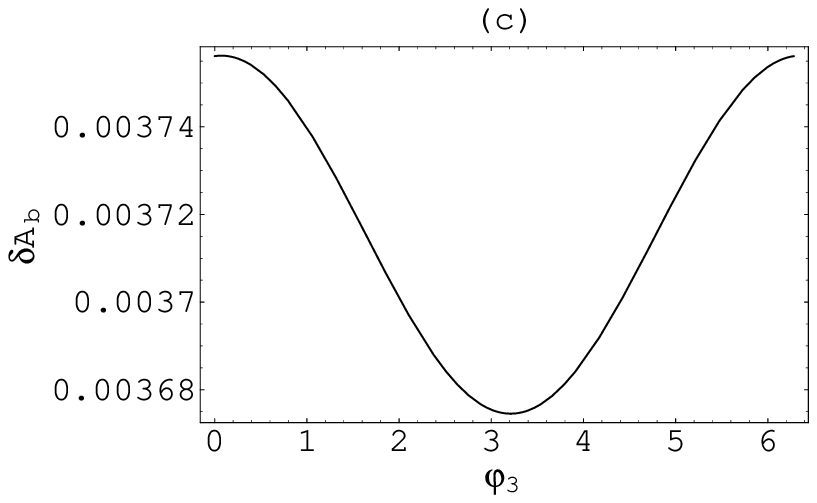}} &
{\epsfxsize=7.cm \epsfbox{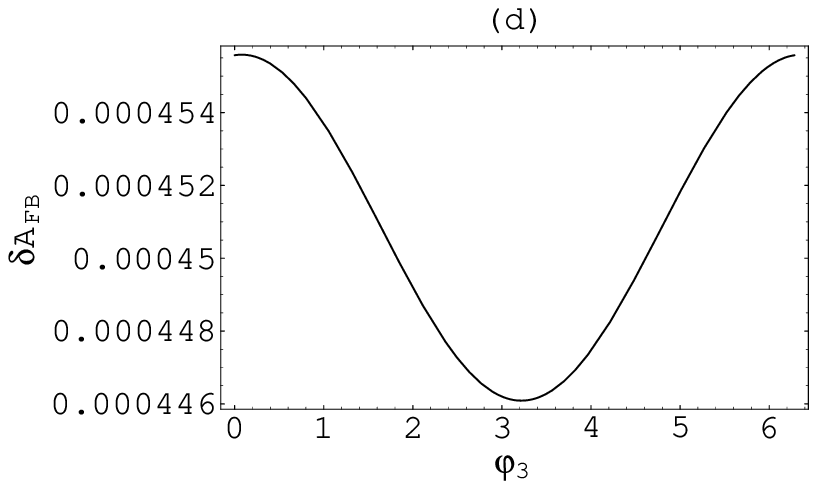}}
\end{tabular}
}
\caption{$(a)$ $\de R_c$, $(b)$ $\de R_l$, $(c)$ $\de A_b$ and $(d)$ $\de A_{FB}^b$
as a function of $\varphi_3$. See the text for the choice of other fixed 
parameters}
\label{other-phi3}
\end{figure}

The change in $Zb\ovl{b}$ also affects other Z-pole observables. 
We consider the effects on $R_c,R_l,A_b,A_{FB}^b$. These are related to
$\de R_b$ or form-factors (see (\ref{eq:Zbb})) as follows
\bea
  \de R_c &=&  R_c - R_c^{\rm SM} 
 =-  \frac{R_c^{\rm SM}}{1-R_b^{\rm SM}}\de R_b,
 \nl
  \de R_l &=&  R_l -R_l^{\rm SM} 
 =  \frac{R_l^{\rm SM}}{1-R_b^{\rm SM}} \de R_b,
 \nl
   A_b &=& \frac{\left|g_R^b+{\al_s \over 3 \pi} F_R^V\right|^2
           -\left|g_L^b+{\al_s \over 3 \pi} F_L^V\right|^2}
   {\left|g_R^b+{\al_s \over 3 \pi} F_R^V\right|^2
           +\left|g_L^b+{\al_s \over 3 \pi} F_L^V\right|^2},
 \nl
   A_{FB} &=& {3 \over 4} A_b A_e.
\eea
In Fig.~\ref{other-phi3} we show $\de R_c,\de R_l,\de A_b,\de A_{FB}^b$
as a function of $\phi_3$ fixing $\phi_b=0$. 
The current experimental values are given by~\cite{Drees:2001xw},
\bea
  R_c &=& 0.1719 \pm 0.0031, \quad
  R_e = 20.804 \pm 0.050, \nl
  A_b &=& 0.922 \pm 0.020, \quad
  A_{FB}^b = 0.0990 \pm 0.0017.
\eea
The deviations from the SM are much smaller than the experimental errors,
and the above Z-pole observables are consistent with the SM precitions.

\section{Conclusions}
The MSSM scenario with light sbottom ($m_{\tilde{b}_1} \sim m_b$) 
and light gluino ($m_{\tilde{g}} \sim 12 -16$ GeV) is quite interesting
and can solve the longstanding problem in the $b$-quark production at
Tevatron. However it receives strong constraint from $Z$-pole precision
tests in the CP conserving scenario.

We considered the effects of CP violating phases $\varphi_3$ and
$\varphi_b$ in the gluino mass
parameter and sbottom mass-squared matrix on Z-pole observables. 
We showed that the 
strong constraint on $R_b$ is significantly relaxed 
in this CP violating MSSM.

{\it Note Added:} After finishing this paper, we received a paper~\cite{Cho:2002mt}
considering the constraint by gauge boson propagator on the MSSM scenario
with light sbottom and gluino,
but in the CP conserving case. Hence it is orthogonal to our work.

\acknowledgements
We thank P. Ko for useful comments.

\newpage

\psfrag{rbbb}[][][0.85]{{\Large $\de R_b [10^{-3}]\hspace{2cm}$ }}
\psfrag{pbbb}[][][0.85]{{\Large $\varphi_b$}}
\psfrag{p333}[][][0.85]{{\Large $\varphi_3$}}

\psfrag{mmmsssbbb}[][][0.85]{{\Large $m_{\tilde{b}_2}$ [GeV]}}
\psfrag{dddrrrbbb}[][][0.85]{{\Large $\de R_b [10^{-3}]$}}

\vfil\eject
\end{document}